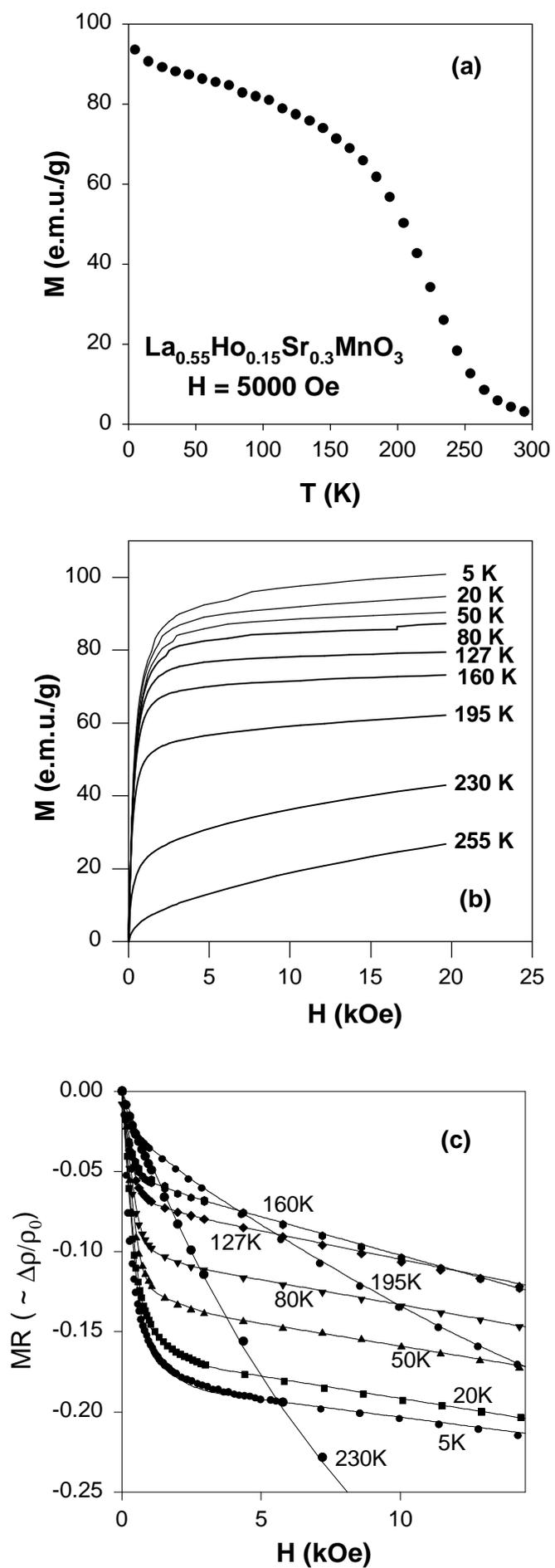

Figure 1 (P Raychaudhuri et al)

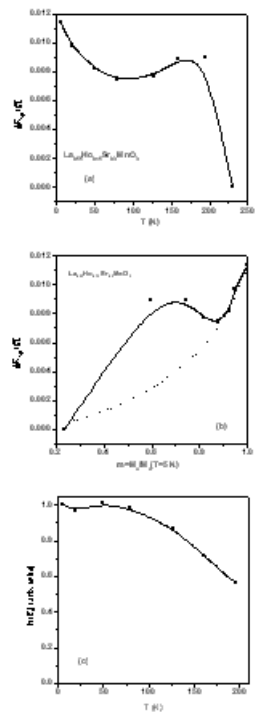

Figure 2 (P Raychaudhuri et al)

# The effect of the total density of states at Fermi level on spin polarised tunnelling in granular $La_{0.55}Ho_{0.15}Sr_{0.3}MnO_3$

**P. Raychaudhuri[†], A. K. Nigam and R. Pinto**
*Tata Institute of Fundamental Research, Homi Bhabha Rd, Mumbai-400005, India.*

and

**Sujeet Chaudhary and S. B. Roy**
*Centre for Advanced Technology, Indore-452013, India.*

*Abstract:* We study the spin polarised tunnelling mechanism through magnetisation and magnetoresistance in the granular polycrystalline colossal magnetoresistive manganite, $La_{0.55}Ho_{0.15}Sr_{0.3}MnO_3$. This system has a ferromagnetic transition temperature ($T_c$) of 255 K associated with a metal-insulator transition around the same temperature. We have investigated dependence of the magnetoresistance due to spin polarised tunnelling on temperature and reduced magnetisation of the sample. We discuss the significance of our results within the realm of a model recently proposed by us to explain the spin polarised tunnelling in granular $La_{0.7}Sr_{0.3}MnO_3$, in the light of the recent finding by A. Biswas et al. (cond-mat/9806084) regarding the evolution of the total density of states at Fermi level as a function of temperature in colossal magnetoresistive materials.

[†] e-mail: pratap@tifr.res.in

Granular itinerant ferromagnets have attracted considerable attention in recent times due to the large magnetoresistance observed in these materials in the polycrystalline form[1,2,3,4]. This extrinsic magnetoresistance which is absent in single crystals under similar conditions arises from spin polarised transport at grain boundaries. This phenomenon is important both from the fundamental and technological point of view since it could be exploited to find new schemes to synthesise materials with giant magnetoresistance.

Recently we have proposed a theoretical model based on intergranular spin polarised tunnelling[5] to explain this extrinsic magnetoresistance in polycrystalline $La_{0.7}Sr_{0.3}MnO_3$. In that model we assumed that the evolution of up and down density of states (DOS) in the material to be reasonably well described by the ferromagnetic Kondo Hamiltonian[6,7,8,9]

$$H = -t \sum_{\langle i,j \rangle} c_{i\sigma}^{\dagger} c_{j\sigma} - J_H \sum_i \boldsymbol{\sigma}_i \cdot \mathbf{s}_i, \quad ----(1)$$

where $t$ is the nearest neighbor hopping energy of the $e_g$ electron and $J_H$ is the local ferromagnetic Hund's rule coupling between the $e_g$ electron spin $\mathbf{s}_i$ and the $t_{2g}$ spin $\sigma_i$ at the i-th site. According to that model the tunnelling conductivity between two grains is given by the expression

$$\sigma(\theta) \propto (1/2)(n_\uparrow + n_\downarrow)^2 [1 + m^2 \cos\theta], \quad ----(2)$$

where $n_\uparrow$ and $n_\downarrow$ are the respective up and down DOS at the Fermi level, m is the reduced spontaneous magnetisation $M_s(T)/M_s(T \rightarrow 0)$, and $\theta$ is the angle between the magnetisation of the two grains. In a field larger than the technical saturation field of the ferromagnet $\theta \rightarrow 0$ thus giving rise to a total resistance drop

$$\Delta R_{spt} \propto (1/\sigma(\theta) - 1/\sigma(0)). \quad -----(3)$$

In a real system with many grains and grain boundaries $\cos\theta$ is replaced by $\langle\cos\theta\rangle$ where the average is over many grain boundaries. In ref. 5 we had shown that this expression fits well in $La_{0.7}Sr_{0.3}MnO_3$ up to 300 K. However, there the implicit assumption was the total DOS at the Fermi level, namely $n_\uparrow+n_\downarrow=N(E_F)$ remains a constant over the whole temperature range. This assumption though valid in the ferromagnetic regime is expected to fail in colossal magnetoresistive manganites at temperatures close to $T_c$, since the ferromagnetic transition in these systems is normally associated with a metal-insulator transition. Thus close to $T_c$ one expects a rapid change in the DOS at Fermi level. This has recently been confirmed from tunnelling spectroscopy studies on a variety of manganites by A. Biswas et al.[10]. They observe the DOS to be almost constant as a function temperatures up to around $0.8T_c$ after which it drops off rapidly and vanishes just before $T_c$. Thus in order to study the effect of $N(E_F)$ on spin polarised tunnelling one has to investigate up to temperatures close to $T_c$.

In this paper, we study the spin polarised tunnelling in $La_{0.55}Ho_{0.15}Sr_{0.3}MnO_3$ from 5 K to $T_c$(~255 K). The polycrystalline samples (2mm×1.5mm×15mm) were prepared through conventional solid state reaction route starting from $La_2O_3$, $Ho_2O_3$, $SrCO_3$ and $MnO_2$. Details regarding sample preparation have been described elsewhere[11]. Magnetisation measurements were done on a Quantum Design SQUID magnetometer. The magnetoresistance (MR) was measured by conventional 4-probe technique using a home made superconducting magnet. The addition of small amount of holmium brings down the $T_c$ to suit the attainable temperature ranges in conventional low temperature cryostats.

Figure 1(a) shows the magnetisation versus temperature (M-T) at 5000 Oe from 5 K to 300 K. The $T_c$, determined from the maximum in the double derivative of the M-T curve was 255 K. Figure 1(b) shows the magnetisation versus field (M-H) at various temperatures below $T_c$ up to a field of 20 kOe. The spontaneous magnetisation ($M_s$) was determined by extrapolating back to zero the linear high field region of the M-H curve at low temperatures and from the Arrot plot at temperatures close to $T_c$. Figure 1(c) shows the MR ($\sim(\rho(H)-\rho(0))/\rho(0)\sim\Delta\rho/\rho_0$) versus field at various temperatures. To separate out the contribution due to spin polarised tunnelling we used a scheme developed earlier[12]. It was shown that the curve fitted well with an expression of the form

$$\text{MR} = -A' \int_0^H f(k)dk - J\text{H} - K\text{H}^3 \text{ with } f(k) = A\exp(-Bk^2) + Ck^2\exp(-Dk^2), \text{ --(4)}$$

where the first term gives the contribution from spin polarised tunnelling and the other two terms give the intrinsic contribution from Zener double exchange. The solid lines in figure 1(c) shows the fitted curve using this expression. The total resistance drop due to spin polarised tunnelling ($\Delta R_{spt}$) is given by $R_0 A' \int_0^H f(k)dk$, where $R_0$ is the zero field resistance and the parameters A,B,C,D are taken from the fitted curve at that temperature. One interesting point to note is that we did not observe any contribution to the MR from spin polarised tunnelling at 230 K though the $T_c$ is around 255 K. The significance of this observation in context of the observations made by A. Biswas et al. will be discussed later. Figure 2(a) shows $\Delta R_{spt}$ as a function of temperature. One normally expects this quantity to drop monotonically with temperature due to decreasing polarisation of the conduction band. However at temperatures close to $T_c$, $\Delta R_{spt}$ shows an increase reaching a maximum before dropping to zero. The same feature is observed when $\Delta R_{spt}$ is plotted as a function of m (figure 2(b)), where $\Delta R_{spt}$ does not decrease monotonically as m is reduced but a broad hump is observed.

To understand the above results we consider equations 2 and 3. First on physical grounds we argue that $\langle\cos\theta\rangle < 0$. There are two interactions which govern the direction of magnetisation in the polycrystalline grains of a ferromagnet: the easy axes which tend to align the magnetisation randomly in different grains and the dipolar interaction which tend to favour antiparallel alignment between grains. Thus $180^0 > \langle\theta\rangle > 90^0$ giving $-1 \ll \langle\cos\theta\rangle < 0$. The dotted line in figure 2(b) shows the typical nature of $\Delta R_{spt}$ as a function m of derived from equation 2 and 3 with $\langle\cos\theta\rangle = -0.63$ when $N(E_F)$ is constant. This curve fits well at low temperatures. However, from equations 2 and 3 we see that the expression of $\Delta R_{spt}$ has a term $\{N(E_F)\}^2$ in the denominator. Thus as one approaches $T_c$, $\Delta R_{spt}$ is enhanced due to the decrease in $N(E_F)^{10}$. We have attempted to extract the temperature variation of $N(E_F)$ by calculating the quantity $\sqrt{\Delta R_{spt}(theoretical)/\Delta R_{spt}(experimental)}$, where $\Delta R_{spt}(theoretical)$ is taken from the computed curve. The calculated quantity is shown in figure 2(c). $N(E_F)$ is constant up to about 100 K and then shows a trend to decrease. However, there could be a large error owing from the fact that the range over which the theoretical curve can be fitted is very small. Closer data near $T_c$ would actually be useful to probe the effect of $N(E_F)$ on $\Delta R_{spt}$.

The other interesting feature of this study is the vanishing of $\Delta R_{spt}$ at 230 K. It has been shown through tunnelling studies[10] in $La_{0.7}Pb_{0.3}MnO_3$ that the DOS at Fermi level actually drops to zero before $T_c$. In the vicinity of $T_c$ a gap opens up near $E_F$. At present the details regarding the nature of this gap is not yet known.

In summary, we have shown that spin polarised tunnelling in $La_{0.55}Ho_{0.15}Sr_{0.3}MnO_3$ deviates from the earlier model on spin polarised tunnelling proposed for $La_{0.7}Sr_{0.3}MnO_3$ based on a constant DOS at temperatures close to $T_c$. This deviation can be attributed to the variation in the total density of states (spin up + spin down) at Fermi level with temperature.

***Acknowledgement:*** The authors wish to thank Praveen Chaddah and S. Ramakrishnan for their keen interest in this work. They would also like to thank R. S. Sannabhadti and A. Patade for technical help.

**Figure Captions:**

Figure 1: (a) Magnetisation versus temperature of $La_{0.55}Ho_{0.15}Sr_{0.3}MnO_3$ at 5 kOe. (b) Magnetisation versus field up to 20 kOe at various temperatures. (c) Magnetoresistance as a function of field (dots) and the fitted curves (solid lines) using equation 4 at various temperatures. The typical values of the parameters at 5 K are $A = -0.2678$, $B = 2.1665$, $C = -0.0125$, $D = 0.4300$ and $J = 2.73 \times 10^{-3}$. The value of $K$ is nonzero only close to $T_c$.

Figure 2: (a) $\Delta R_{spt}$ as a function of temperature from 5 K to 230 K; the solid line is a guide to the eye. (b) $\Delta R_{spt}$ as a function of the reduced magnetisation $m = M_s(T)/M_s(5K)$; the dotted line is the theoretical curve for constant $N(E_F)$ using equation 2 and 3 for $\langle \cos\theta \rangle = -0.63$ and the solid line is a guide to the eye. (c) Temperature variation of $N(E_F)$ derived using equation 3.